\documentstyle[12pt]{article}
\begin{document}
\vskip 1.5cm
\begin{flushright}
{\bf Preprint JINR E2-95- 99}
\end{flushright}
\vskip 1.5cm

\newcounter{my}
\newcommand{\he}
   {\stepcounter{equation}\setcounter{my}{\value{equation}}
\setcounter{equation} 0
   \renewcommand\theequation{\arabic{my}\alph{equation}}}
\newcommand{\she}
   {\setcounter{equation}{\value{my}}
\renewcommand\theequation{\arabic{equation}} }
\newcommand{\ds}{\displaystyle}
\newcommand{\la}[1]{\label{#1}}
\newcommand{\re}[1]{\ (\ref{#1})}
\newcommand{\nn}{\nonumber}
\newcommand{\ed}{\end{document}}
\newcommand{\be}{\begin{equation}}
\newcommand{\ee}{\end{equation}}
\newcommand{\ba}{\begin{eqnarray}}
\newcommand{\ea}{\end{eqnarray}}
\newcommand{\baz}{\begin{eqnarray*}}
\newcommand{\eaz}{\end{eqnarray*}}
\newcommand{\bb}{\begin{thebibliography}}
\newcommand{\eb}{\end{thebibliography}}
\newcommand{\ct}[1]{${\cite{#1}}$}
\newcommand{\ctt}[2]{${\cite{#1}-\cite{#2}}$}
\newcommand{\bi}[1]{\bibitem{#1}}
\newcommand{\e}{\varepsilon}
\newcommand{\eee}{\epsilon}
\newcommand{\bI}{\bar I}
\newcommand{\bq}{{\bar q}}

\begin{center}
{\bf Instanton Effects in  $N\bar N$\\
Annihilation} \\[1cm]
{N.I.Kochelev}\\[1cm]
Laboratory of Theoretical Physics,\\
Joint Institute for Nuclear Research\\
Dubna, SU-141980, Moscow region, Russia
\footnote{ E-mail:KOCHELEV@MAIN1.JINR.DUBNA.SU}\\[2cm]
\end{center}
\begin{center}
{\bf Abstract}\\[0.2cm]
\end{center}
It is shown that specific spin-flavour properties of the nonperturbative
interaction between quarks induced by instantons allow  us to
explain the peculiarities of the OZI rule violation in  $N \bar N$
annihilation. New experiments to  test the instanton mechanism of the
OZI rule violation are proposed.
\newpage
\begin{center}
{\bf Introduction}\\[0.2cm]
\end{center}

Recently, in papers \ct{a1} and \ct{a2} it has been mentioned that
the instanton
induced interaction between quarks \ct{a3} can give an essential
contribution to  processes of  $N \bar N $ annihilation.
However, the most fundamental peculiarities of the
OZI rule \ct{a4} violation in these reactions  \ct{a5}-\ct{a29}
have been unanswered.

In this note, a detailed analysis of the available data on the
OZI rule violation in  $N \bar N$ annihilation is made in the framework
of the model in which instantons determine spin effects in hadron spectroscopy
\ct{a30} and hadron reactions at high energies \ct{a31}.
\footnote{ For  introduction to  instanton effects in spin physics
 see  review \ct{a31} and references therein.}.

This model was suggested several years ago and  was very successful in
the description of  hadron properties.
It was shown \ct{a30} that specific spin-flavour properties of
  instanton-induced interaction between quarks
lead to essential violation of the OZI rule in  the
meson pseudoscalar nonet. At the same time, because the interaction
of that kind is absent in the vector nonet, this fact  explains
 the $\omega-\Phi$ mixing angle an being small \ct{a32}, \ct{a33}.
Later, this model was successfully applied   to resolve the so called
"spin crisis" \ct{a34} connected with the anomalous contribution of
 strange quarks to the proton spin and also to explain
the anomalous Gottfried sum rule violation \ct{a35}  caused by
the violation of the $ SU(2)_f $-flavour symmetries of the
nucleon quark sea \ct{a36}.

Achievements of the instanton mechanism in explanations of the
pecularities of the OZI rule violation in meson spectroscopy
and deep-inelastic scattering give  us serious arguments for
the assumption that just this mechanism is responsible for the
violation of the OZI rule in  $N\bar N$ annihilation.

\begin{center}
{\bf Instanton Mechanism for  OZI\\
rule violation in  $N \bar N $ annihilation}
\\[0.2cm]
\end{center}
The value of the OZI rule violation is determined by the ratio of
the matrix-elements \ct{a37}:
\begin{equation}
Z = \frac{ M(A+B\rightarrow \bar s s + X)}
{[M(A+B\rightarrow \bar u u +X) + M(A+B\rightarrow \bar d d +X)]/\sqrt{2}}.
\label{z}
\end{equation}
{}From this one can get the following form for the cross
section ratio:
\begin{equation}
R = \frac{\sigma(A+B \rightarrow\phi X)}{\sigma(A+B\rightarrow\omega X)}
= (\frac{Z+\tan{\delta}}{1-Z\tan{\delta}})\cdot f, \label{R}
\end{equation}
where $\delta = \Theta - \Theta_i $  is the deviation from the
ideal mixing angle  $\Theta_i=35.3^0$ in the vector nonet
 ($\Theta=39^0$ follows from the quadratic
 Gell-Mann-Okubo mass formula and
$\Theta=36^0 $ comes from the linear formula)
 and $f$ is a phase space factor.

In Table 1 (see \ct{a37}) the experimental data on the OZI violation for
different final hadronic states    recently obtained in the $ N\bar N$
annihilation at rest are presented \ct{a5}-\ct{a16}.
The main conclusion which can be made from the analysis of these
data is  the value of the OZI violation  is  extremely
sensitive to values of the spin of  initial and final particles.
So, the most  remarkable fact is that the large violation comes only
from the $S$-wave state of initial nucleons.  At the same
time the violation in  the $P$-wave state is  small.
 This fact is very difficult to
explain in the framework of the conventional mechanism of the
OZI violation through the $K-$mesons rescattering in the intermediate state
\ct{a38} (see the discussion in \ct{a37}).

It also follows from Table 1 that the value of the violation
is very sensitive to  quantum numbers of  final particles.
So, the essential violation is observed in the reactions
$N\bar N\rightarrow \Phi\gamma$ and $N\bar N\rightarrow \Phi\pi$.
At the same time the violation in the reactions $N\bar N\rightarrow \Phi\rho$,
$N\bar N\rightarrow \Phi\pi\pi$, and $N\bar N\rightarrow\Phi \eta$
is almost absent.

A strong  dependence on the spin and flavour of interacting particles
 is a fundamental peculiarity of the
quarks interaction  induced by instantons \ct{a3}, \ct{a39}
\footnote{The Lagrangian \re{Ins1} was
obtained under the assumption $p\rho \ll 1 $,
 where $p$ is  the characteristic
momentum of  quarks. Taking into account  the inequality of
$p\neq 0 $  one obtains   some form factor in \re{Ins1} which depends
 on the virtualities of  quarks \ct{a40}.}:

\begin{eqnarray}
{\cal L}_{eff}^{(N_f=3)}&=&\int d\rho n(\rho)\left\{\prod_{i=u,d,s}\relax
(m_i\rho-\frac{4\pi}{3}\rho^3\bar q_{iR}q_{iL})+\nonumber\right.\\
&{\ }&\frac{3}{32}(\frac{4}{3}\pi^2\rho^3)^2
[(j_u^aj_d^a-\frac{3}{4}j_{u\mu\nu}^a\relax
j_{d\mu\nu}^a)(m_s\rho-\frac{4}{3}\pi^2\rho^3
\bar q_{sR}q_{sL})+\nonumber\\ \relax
&{\ }& \frac{9}{40}(\frac{4}{3}\pi^2
\rho^3)^2d^{abc}j_{u\mu\nu}^aj_{d\mu\nu}\relax
j_s^c+2perm.]+\frac{9}{320}(\frac{4}{3}\pi^2\rho^3)^3
d^{abc}j_u^aj_d^b\relax
j_s^c+\nonumber\\  \relax
&{\ }&\frac{igf^{abc}}{256}(\frac{4}{3}\pi^2\rho^3)^3
j_{u\mu\nu}^aj_{d\nu\lambda}^b\relax
j_{s\lambda\mu}^c+(R\longleftrightarrow L)\left.\right\} ,
\label{Ins1}
\end{eqnarray}
where $q_{R,L}={(1\pm\gamma_5)\over2}q(x),
\ j_i^a=\bar q_{iR}\lambda^aq_{iL},{\ }
j_{i\mu\nu}^a=\bar q_{iR}\sigma_{\mu\nu}\lambda^aq_{iL}$, and   $n(\rho)$
is the instanton density.
Neglecting quark masses and using  the Fierz transformation one can
transform \re{Ins1}  to the following flavour determinant:
\[
{\cal L }_{eff}^{(N_f=3)} \sim \e^{ijk}\e_{i^\prime j^\prime k^\prime}
\times
\left[
\left(\bq_{Ri_1}q_L^{i^\prime}\right)
\left(\bq_{Rj}q_L^{j^\prime}\right)
\left(\bq_{Rk}q_L^{k^\prime}\right)
\right. +
\]
\be
\left.
+\frac{3}{8(N_c+2)}
\left(\bq_{Ri}q_L^{i^\prime}\right)
\left(\bq_{Rj}\sigma_{\mu\nu}q_L^{j^\prime}\right)
\left(\bq_{Rk}\sigma_{\mu\nu}q_L^{k^\prime}\right)
\right]
\la{Ins2}
\ee

One can emphasize three the most remarkable characteristics of this
interaction.  First, it contributes only to the
$ S-$ wave scattering of  initial particles. This property
comes from $O(4)$ symmetry of the instanton solution \ct{a41},
which leads to the point-like behavior of the quarks scattering
 amplitude off instanton (see for example \ct{a42}).

Second, helicities of incoming and outcoming quarks
from instanton (antiinstanton) are strongly  correlated:
all incoming quarks are left (right), all outcoming quarks are
right (left). Specific helicity properties of the t'Hooft's
interaction \re{Ins1} are determined by the helicity structure
of the zero fermion modes in the instanton field which give  the
dominated  contribution to the instanton
induced quark scattering  amplitude.

The most outstanding consequence of this structure is a very large
violation of the quark helicities in the instanton field:
\be
\Delta Q_5=-2N_f .
\label{he}
\ee
This fact provides  the basis for explanation  of the
"spin crisis"   in the framework of the instanton mechanism
(see a discussion in \ct{a31}).
It should be mentioned also that just this property  of the quark and
lepton interaction off the electroweak
instantons
violates the baryon number conservation \ct{a3}. At present, the
possibility of its anomalous violation due to a multiple creation of the
gauge bosons from the instanton vertex at high energies is widely discussed
 (see \ct{a42} and references therein).

The third property of the instanton Lagrangians \re{Ins1}, which has
a direct relation to the OZI rule violation is that it is not equal to
zero only
for different  flavours of the interacting quarks.
 Just this peculiarity of the Lagrangian
\re{Ins1}
 enhances  the probability of  transitions
 between different quark
flavours and determines  the mixing angles of the $SU(3)_f$-multiplets
\ct{a30}.

On the whole, all these peculiarities are just the reason
 for splitting
of masses between hadron multiplets  ($\pi-\rho, N-\Delta, \eta-
\eta^\prime $ and so on). At the same time, the absence of the P-wave
instanton-induced quark interaction leads to the small spin-orbital
splitting in the hadron excited states \ct{a30}. It  provides a
 solution of the   old problem of hadron spectroscopy
connected with  the observed small spin-orbital splitting.
So,  many  models which take into account only a long-range
gluon exchange contribution to the spin-spin quark interaction,
give also a very large spin-orbital splitting which  contradicts
the  experimental data \ct{a43}.

Let us consider now instanton effects in the $\bar NN$ annihilation.
We will suppose that just instantons lead to a large violation of
the OZI rule in these reactions. Other mechanisms of this violation,
 from our point of
view, can be either annihilation through perturbative gluons or
rescattering  \ct{a38}. However, it was shown in papers \ct{a32},
\ct{a33} that the gluon perturbative exchange leads  to
mixing angles in the pseudoscalar nonet  to one order of
 magnitude smaller than instanton
exchange. Moreover, this mechanism predicts the wrong sign of the
$\omega-\Phi$- mixing \ct{a32}. The second mechanism, probably,
is also negligible
when one takes into account the form factors in
the K-meson-nucleon interaction vertex \ct{a441}
\footnote{In paper \ct{a44} it was shown that the K-meson contribution
to the nucleon strange sea should be very small when one takes into account
form factors in the meson-nucleon vertex.}.

Recently, the conception of the "internal polarized strange sea"
(IPSS) was proposed \ct{a37}. Although the fundamental mechanism which
could lead to this phenomenon was not explained in \ct{a37}, this
hypothesis  is supported by the result of
  measurements of the part of the proton spin
carried by the strange quarks by the EMC,
 SMC, E-142, E-143 Collaborations \ct{a34}. We will compare the predictions
of the IPSS model with predictions of the instanton model under discussion.

By using the specific properties of the instanton-induced Lagrangians
\re{Ins1}, \re{Ins2}, one can formulate some selection rules
in order to predict the values of the OZI rule
violation in  different channels.

First of all, it is expected that  a large OZI rule violation
takes place only
in the $S$- wave initial states, because the interaction \re{Ins1}
is  not equal to zero  for the S-wave quark-quark interactions.
 This rule  is well
 fulfilled for  all channels where we have experimental data for different
relative weights of the $S-$ and  $P-$ waves in the initial nucleon state
(see Table 1). It should be pointed out that just the absence of the
$P-$ wave interaction induced by the instanton  explains
observed smallness of the spin-orbital splitting in hadron excited
states \ct{a30}.

The second selection rule follows from the spin structure of
different terms in \re{Ins2}. So, the largest
violations in  that kind of reactions, to which  the first term in \re{Ins2}
does contribute, are expected. One can easy understand
that this term corresponds
to the total initial quark spin $S_{q\bq}=0$, and therefore the OZI
rule violation has the maximum magnitude for reactions with $S_{N\bar N}=0$.
This effect is actually observed in the reaction
 $\bar N N\rightarrow\Phi\gamma$.
In hadron spectroscopy, the same effect leads to a very large contribution
of the instanton induced interaction to the masses of particles from
the pseudoscalar nonet and to the dominance of the scalar diquark in the
nucleon wave function \ct{a30}.
{}From this rule it  follows that the value of the violation of the
OZI rule in the reaction $\bar N N\rightarrow \Phi\pi$
should be smaller. This reaction originates through the $^3S_1$ state of two
nucleons, and therefore only the second term in \re{Ins2}, suppressed as
$1/N_c$, gives a nonzero contribution. The dominance of the $^3S_1$ state
in the reaction leads to the specific angular dependence of the $K-$mesons
from $\Phi-$decay
\be
W(\Theta)\approx1-cos^2(\Theta).
\label{ang}
\ee
This dependence has been observed in the experiment \ct{a45}.

It should be emphasized that we do not expect a significant
violation of the OZI rule in the production of the tenzor $2^{++}$
mesons because the instanton vertex \re{Ins2} does not include
the term with appropriate quantum numbers.

Further, the third selection rule results from
 a suppression of the direct creation of the vector mesons
in the instanton field. This suppression
 comes from specific helicity properties of the
quark zero modes. So, for  quark on zero mode the following
condition should be satisfied \ct{a3}:
\be
\vec \sigma_q\oplus\vec c_q =0,
\la{zero}
\ee
where $\vec \sigma_q$ is a spin and $\vec c_q$ is a colour spin
for the $SU(2)_c$ subgroup of the $SU(3)_c$-colour group.
{}From relation \re{zero} and from the fact that quarks
outcoming from the instanton
should  have the same helicities, it  immediately follows  that
without spin-flip the instanton production of the colourless vector
mesons is forbidden. In the one-instanton approximation, the quark
spin-flip can be  induced only by the current quark masses.
Therefore, the production  of the vector mesons
which consist of the light $u$- and $d$-quarks should be small.
Just this property leads to a very small OZI rule violation
in the reactions $\bar N N\rightarrow\Phi \rho$,
 $\bar N N\rightarrow\Phi\omega$. In the quark model
 \ct{a30} it results in the smallness of
the $\Phi-\omega$-mixing angle.

The forth selection rule is connected with the flavour structure of
the Lagrangians \re{Ins1}, \re{Ins2}. So, it is not equal to zero
only for different quark flavours. Therefore, the violation of the
OZI rule in the reactions, where the final hadrons include the same kind
of  quarks, for example in reactions $\bar N N\rightarrow\Phi 2\pi$ and
$\bar N N\rightarrow\Phi\Phi$, should be suppressed by the parameter
of the instanton density in the QCD vacuum \ct{a33}:
\be
f=\pi^2\rho^4n_I\approx\frac{1}{20}.
\nn
\ee
This rule also  decreases  the OZI rule violation
in the reactions $\bar N N\rightarrow\Phi\eta$ and
$ \bar N N\rightarrow\Phi\eta^\prime $ because the wave functions of
$\eta$ and $\eta^\prime$-mesons include some part of the strange
quarks. However, in this case, this explanation
is not enough because the mesons can be created through their nonstrange
isospin singlet component
\footnote{The author is grateful to S.B. Gerasimov for useful
discussions of this problem.}. From our point of view,
the suppression of the yield of the
$\eta$ and $\eta^\prime$-mesons in the  reactions
 $\bar N N\rightarrow\Phi\eta$ and $ \bar N N\rightarrow\Phi\eta^\prime $
is directly related to the "spin crisis" \ct{a34}. In  papers
\ct{a46} it was shown that the EMC result, can be understood as
decoupling of the isosinglet $\eta^\prime_0$ from the nucleon.
In the framework of the instanton model, the decoupling of the isosinglet
meson comes from the fact that the instanton-induced interaction \re{Ins1}
violates the $U_A(1)$-symmetry in QCD. So, the instanton-induced
interaction is repulsive in
the isosinglet channel \ct{a30}, \ct{a33}, \ct{a47}.
Without taking into  account the octet-singlet mixing, it leads to the
unbound $\eta^\prime_o$-meson state \ct{a47},  and therefore its
interaction with the nucleon should be very small.

The fifth rule comes from the helicity properties of the instanton
vertex. So, in the c.m. frame, all quarks
 incoming into the instanton
should  have the same helicity. It  leads to the enhancement of the
OZI sum rule violation from the $S_{\bar N N}^Z=0$ two-nucleon state,
where $S_{\bar N N}^Z$ is the projection of the total spin of the
$\bar NN$ pair on the direction of their relative motion.

Thus,  some rules, which come from the specific
properties of the instanton-induced interaction between quarks,
have been formulated here.

These rules are  well satisfied  for the available data on
$\bar N N$-annihilation at rest (see Table 1). However, there are the data
on the OZI rule violation in flight experiments \ct{a17}-\ct{a27}.
One of the most remarkable peculiarities of these data is a very fast
decrease of the violation with growing  energy. In the framework of
the instanton model this effect can be   explained easily.
 The instanton is a quasiclassical  object,
extended in the space-time  and therefore all quarks, which
interact with instanton should  have sufficiently small momenta
to provide a large value for the interaction amplitude. It means that the
momenta of the quarks should obey the condition $\mid \vec p\mid
\le1/\rho_c $, where $\rho_c\approx 1.6\  GeV^{-1}$
 is the average instanton size in
QCD vacuum \ct{a33}, and therefore the instanton effects are large only
near the thresholds
\footnote{ It should be mentioned that, probably, the anomalous violation
of the OZI rule is possible also at high energies, but only in events
with a large multiplicity. In these events, large initial energy
transforms to the creation of a large number of  gluons
$(N\sim 1/\alpha_s)$ with small energies $(E\sim1/\rho_c)$.  This
fact provides the anomalous behavior of the spin-dependent structure
function $g_1(x)$ at small $x$ in  QCD and the possibility of the
anomalous violation of the baryon number conservation in the electroweak
theory (see a discussion in \ct{a49}).}.

The instanton model also predicts a strong dependence of  relative
values of the OZI rule violation as a function of the  momentum transfer.
So, the average size of the
instanton $\rho_c\approx 1.6\  GeV^{-1}$ ,  determining  the
momentum transfer dependence  for the reactions
 with OZI violation  significantly differs
 from the confinement size $R_{conf}\approx 5 GeV^{-1}$,  which
determines   the momentum transfer dependence of the processes without
OZI rule violation. As a result, we should have a strong momentum transfer
dependence of the
ratio of the cross-sections of these reactions. This effect is obvious
 in  different reactions (see the discussion in \ct{a37}) and
one cannot find its appropriate explanation  in the framework of the
convenient models.

Another challenging problem for the $\bar NN$ annihilation
 models is the backward peak in the
reaction $\bar P P\rightarrow K^+K^-$ \ct{a50}. In our model this
phenomenon can be explained by a large changing of the helicities
$(\Delta\lambda=-6)$ induced by the instanton. So, every quark spin-flip
leads to the factor
\be
M\sim \bar q_Rq_L\sim sin(\frac{\Theta}{2})
\ee
in the matrix element of the reaction.
 Therefore, one can estimate that the
spin-flip leads to the following angular dependence
of the cross-section:
\be
\frac{d\sigma^{\bar PP\rightarrow K^+K^-}}{d\Omega}\sim
sin^6(\frac{\Theta}{2}).
\ee

Thus, the instanton induced-interaction allows  us to describe,
at least qualitatively, principal  peculiarities of the OZI
rule violation in $\bar NN$-annihilation.
\begin{center}
{\bf Possible Tests of the Model}
\\[0.2cm]
\end{center}

A very important task is to perform  the  experiments, which
would  give a direct indication of the instanton mechanism of the
$\bar NN $-annihilation. One of these important experiments
 is a measurement of the cross section of the
reaction $\bar N N\rightarrow\Phi\gamma$ on a gas target.
This experiment will  test the domination of the $S-$wave
mechanism of the OZI rule violation. So, we predict a sharp  decrease
of the ratio $(\Phi\gamma/\omega\gamma)$ on a gas target, where the
annihilation dominates in the $P-$wave state in comparison
with the result in a liquid
\ct{a6},  where the annihilation  dominates in the $S-$wave  state.
It should be mentioned that the  IPSS model \ct{a37} predicts a
sharp  increase of the ratio, and therefore this experiment
will allow  unambiguous conclusion about
 the reliability of  one of these models.

Another  experiment could be a measurement of
the spin correlations in the reaction \ct{a37}:
\be
\bar P + P\rightarrow K^* + \bar K^*.
\ee
The IPSS model predicts   a strong correlation in the final
$S_{K^*\bar K^*}=2$ state for this reaction. Our model gives the correlation
in the
$S_{K^*\bar K^*}=0$ state because the initial state  $S_{\bar pp}=0$
dominats.

A possible test for the model is the angular distribution of
dileptons in the reaction:
\ba
p + p &\longrightarrow &\phi + X,\\
& &\hookrightarrow e^{+}e^{-} \label{dec}
\ea
So, the IPSS model predicts:
\be
W(\Theta)\approx 1+cos^2(\Theta).
\ee
At the same time, in our model the dependence
\be
W(\Theta)\approx 1-cos^2(\Theta)
\ee
is expected. This form
 results from the longitudinal polarization of the $\Phi$-
meson. It was mentioned above that
the quark-antiquark pair created by the instanton (antinstanton) has
the helicity $\lambda=\pm1$, and it is necessary to flip a helicity of
one of the quarks to create a vector meson. As result we have
a longitudinal polarization of the created vector mesons.

It would be interesting to measure the dependence
\be
R(n)=\frac{\sigma(\bar NN\rightarrow\Phi(n\pi))}{\sigma(\bar NN\rightarrow
\omega(n\pi))}.
\ee
It was pointed out above (see also \ct{a49}, \ct{a51}) that  the anomalous
behavior of the spin-dependent structure function $g_1(x)$ \ct{a34}
at low $x$ is due to the increase of the number of  gluons
from the instanton vertex  created together with a pair of  strange and
nonstrange quarks. It could be shown that the yield of an even number of the
gluons from the instanton vertex is enhanced
(see estimations in \ct{a52}) and therefore after
  hadronization of these gluons to the pions one
 can expect  the oscillator-like behavior of the function $R(n)$ with
maxima at odd numbers of the pions.

{}From our point of view, the  most direct experiments to check the instanton
model are experiments on measurement of the OZI rule violation
with polarized beams. One of these processes could be the reaction
\be
\vec P + \vec P\rightarrow P+P+\Phi.
\ee
 Taking into account  the fact that
 the instanton-induced interaction \re{Ins1} is the $S-$wave interaction,
 we  expect the  enhancement
of  production of  $\Phi$  in the $S_{PP}=0$ state and, respectively,
the following value of two-spin asymmetry in this reaction at rest:
\be
A=\frac{Y_\Phi(\uparrow\uparrow)-Y_\Phi(\uparrow\downarrow)}
{Y_\Phi(\uparrow\uparrow)+ Y_\Phi(\uparrow\downarrow)}\approx-1.
\ee
This experiment is planned \ct{a53}, and its realization would
be very interesting.

\begin{center}
{\bf Conclusion}
\\[0.2cm]
\end{center}
The complex structure of the QCD vacuum caused by
 of the strong nonperturbative fluctuations of the gluon fields -
instantons in the QCD vacuum  manifests nontrivially in  $\bar NN $-
annihilation.

In this note, we have argued that the unique spin-flavour
properties of the instanton-induced interaction between quarks allow
 us to explain the peculiarities of the OZI rule violation
in these reactions.

Further experimental and theoretical investigation of the
instanton effects in the OZI rule violation is very important to shed
light on
the role of the fundamental structure of the QCD vacuum in
$\bar NN $-annihilation.

The author is grateful to P.N.Bogolubov,
A.E.Dorokhov, S.B.Gerasimov, F.Lehar,
O.V.Teryaev, and especially to M.G.Sapozhnikov for useful discussions.

\newpage
TABLE 1. The ratios $R=\phi X/\omega X $ for production of
the $\phi$ and $\omega$ - mesons in
antinucleon annihilation at rest.
 The data are given for annihilation in
liquid hydrogen target (percentage of annihilation from P-wave
is $\sim 10-20 \%$),
gas target ($\sim$61\% P-wave) and
 LX-trigger \ct{a5} ($\sim$86-91\% P-wave).
 \\~\\
\begin{tabular}{llllll}
\hline
Final state& Initial states & B.R.$\cdot10^{4}$& $R\cdot 10^{3}$
&$\left| Z\right|~(\%)$ &Comments\\
\hline
$\phi\gamma$ & $^1S_0,^3P_J$ & $0.17\pm0.04$  & $250\pm89$
&  $42\pm8$ & liquid,\ct{a6}\\
\hline
$\phi\pi^0$ & $^3S_1,^1P_1$ &$5.5\pm0.7$  & $96\pm15$
&  $24\pm2$ & liquid,\ct{a6}\\
$\phi\pi^0$ & &$1.9\pm0.5$ &
&   & gas, \ct{a5}\\
$\phi\pi^0$ & &$0.3\pm0.3$ &
&   & LX-trigger, \ct{a5}\\
\hline
$\phi\pi^-$ &$^3S_1,^1P_1$ &$9.0\pm1.1$
& $83\pm25$
&  $22\pm4$ & liquid,\ct{a9}-\ct{a12}\\
$\phi\pi^-$ & &$14.8\pm1.1$  & $133\pm26$
&  $29\pm3$ & $\bar p d,
p<200~ MeV/c$, \ct{a7} \\
$\phi\pi^-$ & & & $113\pm30$
&  $27\pm4$ & $\bar p d,p>400~ MeV/c$, \ct{a7} \\
$\phi\pi^+$ &  & & $110\pm15$
&  $26\pm2$ & $\bar n p$, \ct{a7} \\
\hline
$\phi\eta$ &$^3S_1,^1P_1$ & $0.9\pm0.3$ & $6.0\pm2.0$
&  $1.3\pm1.2$ & liquid,\ct{a6}\\
$\phi\eta$ & &$0.37\pm0.09$ &
&  & gas, \ct{a5}\\
$\phi\eta$ & &$0.41\pm0.16$ &
&   & LX-trigger, \ct{a5}\\
\hline
$\phi\rho$ & $^1S_0,^3P_J$ & $3.4\pm0.8$ & $6.3\pm1.6$
&  $1.4\pm1.0$ & gas, \ct{a5},\ct{a15}\\
$\phi\rho$ & &$4.4\pm1.2$ & $7.5\pm2.4$
&  $2.1\pm1.2$ & LX-trigger, \ct{a5},\ct{a15}\\
\hline
$\phi\omega$ & $^1S_0,^3P_{0,2}$ &$6.3\pm2.3$  & $19\pm7$
& $7\pm4$ & liquid, \ct{a14},\ct{a16}\\
$\phi\omega$ & &$3.0\pm1.1$  &
&          & gas, \ct{a5}\\
$\phi\omega$ & &$4.2\pm1.4$  &
&          & LX-trigger, \ct{a5}\\
\hline
$\phi\pi^0\pi^0$ &$^{1,3}S_{0,1},^{1,3}P_J$ &$1.2\pm0.6$  & $6.0\pm3.0$
& $1.3\pm2.0$ & liquid,\ct{a6}\\
$\phi\pi^-\pi^+$ & &$4.6\pm0.9$  & $7.0\pm1.4$
& $1.9\pm0.8$ & liquid,\ct{a13}\\
$\phi X,
X=\pi^+\pi^-, \rho$ & &$5.4\pm1.0$  &$7.9\pm1.7$
&$2.4\pm1.0$  & gas, \ct{a5},\ct{a15}\\
$\phi X,
X=\pi^+\pi^-, \rho$ & &$7.7\pm1.7$  &$11.0\pm3.0$
&$4.0\pm1.4$  & LX-trigger, \ct{a5},\ct{a15}\\
\hline\\~\\
\end{tabular}

\end{document}